\documentclass[manuscript]{aastex}

\shorttitle{CME flare relation}
\shortauthors{Bein et al.}

\usepackage{natbib}
\usepackage{graphicx}
\usepackage{amsmath}
\usepackage{rotating}

\begin{document}
\title{Impulsive acceleration of coronal mass ejections: \\ II. Relation to SXR flares and filament eruptions}

\author{B. M. Bein\altaffilmark{1}, S. Berkebile-Stoiser\altaffilmark{1}, A. M. Veronig\altaffilmark{1}, M. Temmer\altaffilmark{1}}
\affil{Kanzelh{\"o}he Observatory-IGAM, Institute of Physics, University of Graz, Universit{\"a}tsplatz 5, 8010 Graz, Austria}

\and

\author{B. Vr{\v s}nak\altaffilmark{2}}
\affil{Hvar Observatory, Faculty of Geodesy, University of Zagreb, Ka{\v c}i\'{c}eva 26, HR-10000 Zagreb, Croatia}

\date{Received/Accepted}

\begin{abstract}
Using high time cadence images from the STEREO EUVI, COR1 and COR2 instruments, we derived detailed kinematics of the main acceleration stage for a sample of 95 CMEs in comparison with associated flares and filament eruptions. We found that CMEs associated with flares reveal on average significantly higher peak accelerations and lower acceleration phase durations, initiation heights and heights, at which they reach their peak velocities and peak accelerations. This means that CMEs that are associated with flares are characterized by higher and more impulsive accelerations and originate from lower in the corona where the magnetic field is stronger. For CMEs that are associated with filament eruptions we found only for the CME peak acceleration significantly lower values than for events which were not associated with filament eruptions. The flare rise time was found to be positively correlated with the CME acceleration duration, and negatively correlated with the CME peak acceleration. For the majority of the events the CME acceleration starts before the flare onset (for 75\% of the events) and the CME accleration ends after the SXR peak time (for 77\% of the events). In $\sim$60\% of the events, the time difference between the peak time of the flare SXR flux derivative and the peak time of the CME acceleration is smaller than $\pm$5 min, which hints at a feedback relationship between the CME acceleration and the energy release in the associated flare due to magnetic reconnection. 

\end{abstract}

\keywords{Methods: statistical, Sun: coronal mass ejections (CMEs), Sun: flares}

\maketitle

\section{Introduction}
  
Solar flares and coronal mass ejections (CMEs) are the two most energetic phenomena in our solar system. Solar flares are abrupt releases of energy up to 10$^{25}$~J within tens of minutes and can be observed in the whole electromagnetic spectrum from radio emission to $\gamma$-rays \citep[e.g.][]{fletcher2011}. CMEs are sporadic ejections of coronal material with velocities in the range of $\sim$100--3000~km~s$^{-1}$ \citep[e.g.][]{yashiro2004, gopalswamy2009}. It is generally accepted that both CMEs and flares are different manifestations caused by magnetic reconnection in the solar corona, but the details how both phenomena are related are still under investigation. 

Various statistical studies using white light coronagraphic observations showed positive correlations between the flare intensity and CME velocity \citep[e.g.][]{moon2002, vrsnak2005, yashiro2009} or CME kinetic energy \citep{burkepile2004, yashiro2009}. The temporal differences between the CME and the flare onset were found to be quite small.  \citet{michalek2009} and \citet{yashiro2009} found a Gaussian distribution for the difference between the flare and CME onsets. According to \citet{michalek2009} the mean difference is 7 min. In these studies, linear back extrapolation of the CME height time curve was used to estimate the CME start time. However, the CME onset times cannot be accurately determined by using only coronagraphic observations which miss the early phase of the CME acceleration and propagation due to the occulter disk.

The CME kinematics typically shows three phases of evolution \citep{zhang2001}. In the initiation phase, the CME rises with velocities of several tens of km~s$^{-1}$, followed by an impulsive acceleration. Thereafter, the CME propagates with almost constant or slowly decreasing/increasing velocity depending on its interaction with the ambient solar wind \citep[e.g.][]{gopalswamy2000}. Recent case studies \citep[e.g.][]{gallagher2003, vrsnak2004, temmer2008} combined EUV images with coronagraphic observations to derive detailed CME acceleration profiles from the CME initiation close to the solar surface until its propagation beyond 15~$R_{\odot}$. In these events the CME acceleration profiles showed a good synchronization with the energy release in the associated flare, as evidenced in the HXR flux or SXR derivative. \citet{maricic2004} studied a set of 22 CME events associated with flares using SOHO EIT, MLSO Mark IV, LASCO C2 and C3. They report correlations between various CME and flare parameters as well as a close synchronization between the CME acceleration and the flare SXR flux derivative for $\geq$ 50\% of the events under study.  

In \citet[][to which we refer to as paper I in the following]{bein2011}, we presented statistics and correlations between various decisive CME parameters  for a sample of 95 events: peak velocity, peak acceleration, acceleration duration, height at peak velocity, height at peak acceleration and initiation height. To this aim we combined EUV images from the Extreme Ultraviolet Imager \citep[EUVI;][]{wuelser2004} with COR1 and COR2 coronagraphic observations onboard the STEREO mission. The high time cadence and the overlapping field of views (FOV) of the different STEREO instruments enabled us to derive detailed and continuous CME height-, velocity- and acceleration-time profiles from the low corona up to about 15~R$_{\odot}$. Out of the 95 CMEs presented in \citet{bein2011}, 70 events could be associated with a GOES (Geostationary Operational Environmental Satellite) flare and 24 events with a filament eruption. 9 events are associated with both, a flare and a filament. In the present paper, we perform a statistical study on the relation between characteristic CME and flare parameters as well as on the temporal relationship between the two phenomena. We also study the characteristic CME parameters separately for events with/without associated flares and with/without associated filament eruptions.

\section{Data}

The CME kinematics was derived from data of the STEREO/SECCHI instrument suite \citep{howard2008}. The STEREO mission consists of two nearly identical spacecraft, STEREO-A, which moves ahead the Earth, and STEREO-B, which moves behind the Earth. The SECCHI package consists of an Extreme Ultraviolet Imager (EUVI), two coronagraphs (COR1 and COR2) and two Heliospheric Imagers (HI1 and HI2). For our study of the CME kinematics we combined data from EUVI, COR1 and COR2.  

EUVI provides observations from the solar chromosphere and low corona in four different wavelengths with a FOV up to 1.7~$R_{\odot}$ \citep{wuelser2004}. For our study we used primarily 171~\AA~observations with a time cadence as high as 75 sec. In some cases it was only possible to track the CME in 195~\AA~observations with a default cadence of 10 min but in some cases as good as 2.5 min. The high time cadence was necessary to derive detailed acceleration profiles of the CME initation and impulsive acceleration phase. Observations in 171~\AA~and 195~\AA~were compared to 304~\AA~filtergrams to check if a filament eruption and/or flare was observed in association with the CME under study.

The inner coronagraph COR1 has a FOV of 1.4 to 4 $R_{\odot}$. For most events it has a nominal time cadence of 5 min but in some cases it was also lower (20 min). The outer coronagraph COR2 has a FOV of 2.5 to 15 $R_{\odot}$ and an observing cadence of 30 min for the polarized brightness images, which were used in our study \citep{howard2008}. The overlapping FOVs between EUVI and COR1 and between COR1 and COR2 made it possible to connect the same CME structure in all three instruments, and thus to study the CME evolution continuously from its initation close to the solar surface up to about 15 $R_{\odot}$. To compare the CME kinematics with the flare SXR evolution we used GOES 1--8~\AA~flux measurements.

\section{Results}

\begin{figure*}
	\centering
		\includegraphics[scale=0.7]{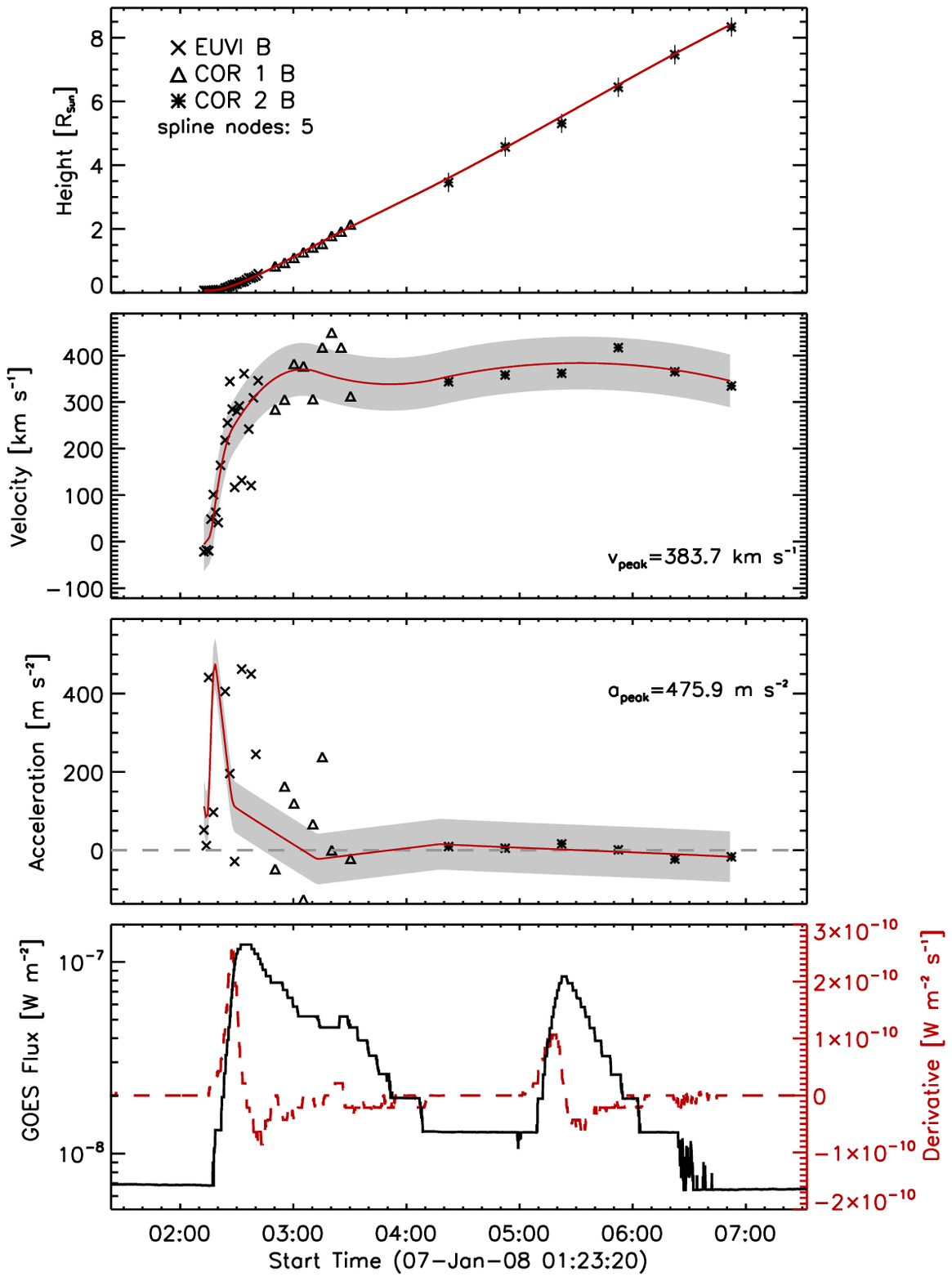}~~~~~~~~~~~~
		\includegraphics[scale=0.7]{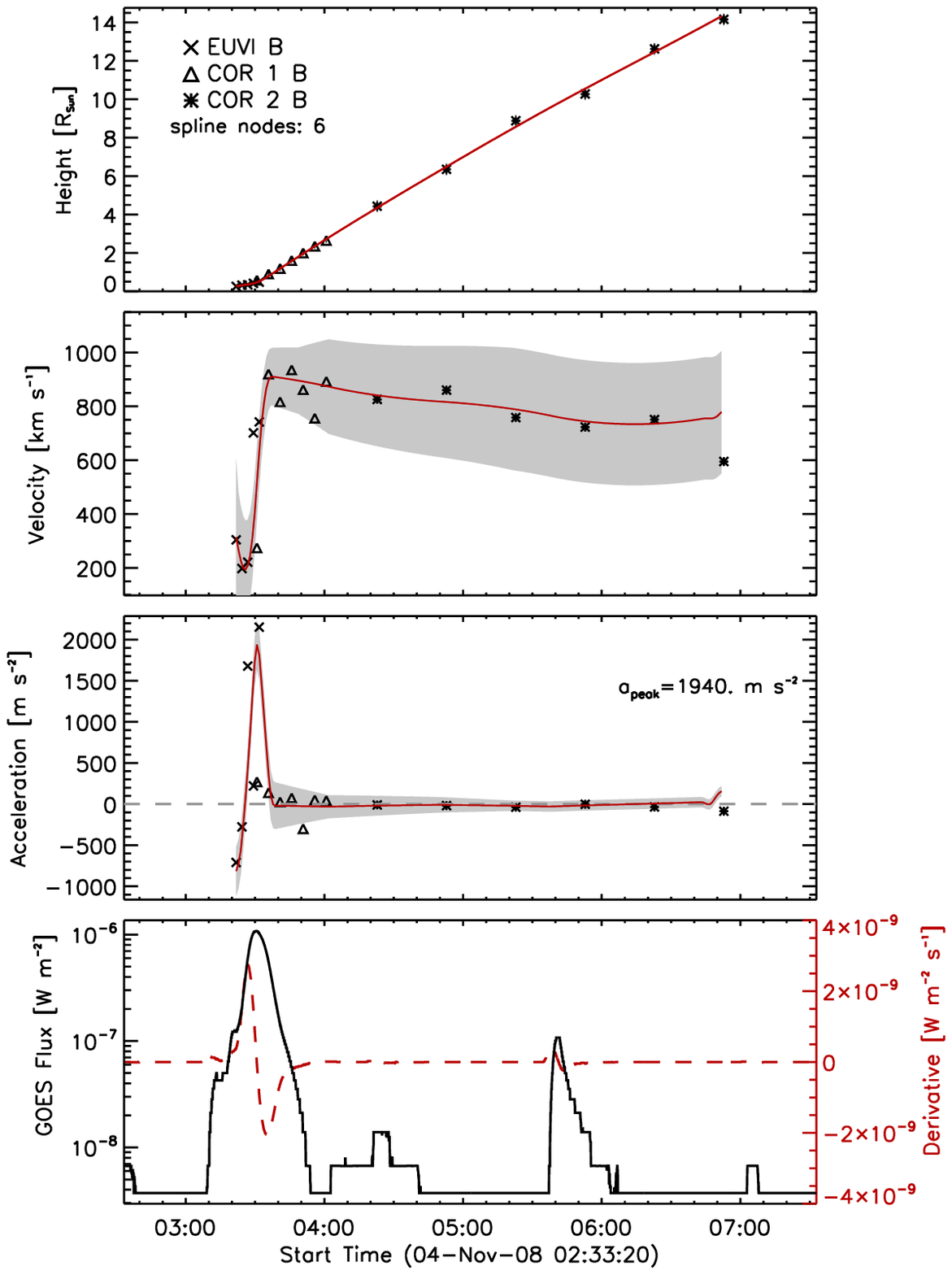}
	\caption{CME kinematics and GOES 1--8 \AA~soft X-ray flux for the CME-flare events that occurred on 2008 January 7 (left) and 2009 January 9 (right). The top panels show the measured CME height-time curve derived from STEREO EUVI (crosses), COR1 (triangles) and COR2 (asterisks) observations. The measurement errors (0.03 $R_{\odot}$ for EUVI, 0.125 $R_{\odot}$ for COR1 and 0.3 $R_{\odot}$ for COR2), which are in some cases smaller than the plot symbols, and the spline fit (solid line) are overplotted. The second and third panels show the CME velocity and acceleration profiles, derived from numerical differentiation of the direct measurements (plot symbols) and the spline fit (solid line) to the height-time curve. The grey shaded area represents the error range of the spline fit. The bottom panels show the GOES flux (black solid line) and its derivative (red dashed line) of the associated flare.}
	\label{height1}
\end{figure*}

\begin{figure*}
	\centering
		\includegraphics[scale=0.7]{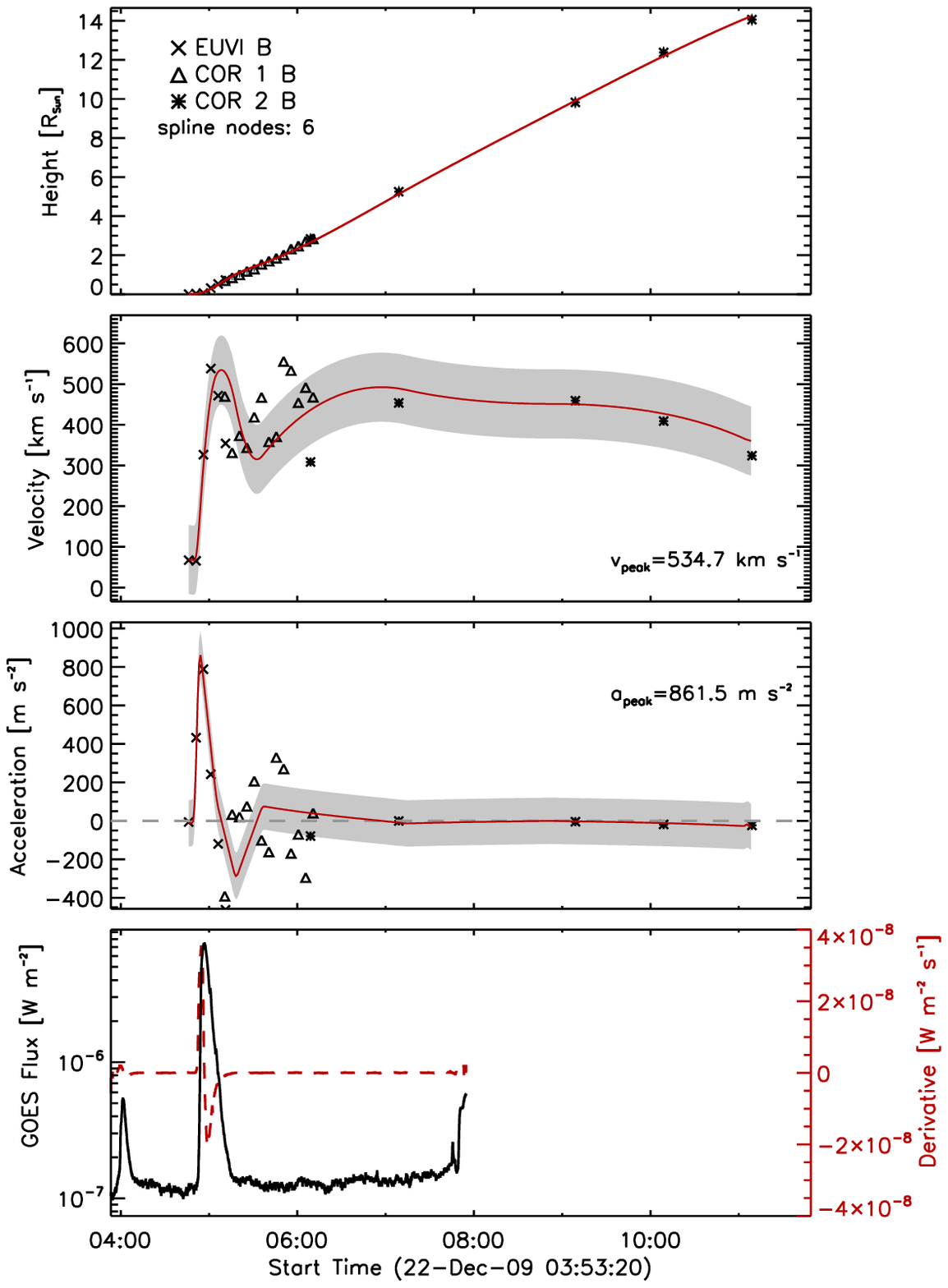}~~~~~~~~~~~~
		\includegraphics[scale=0.7]{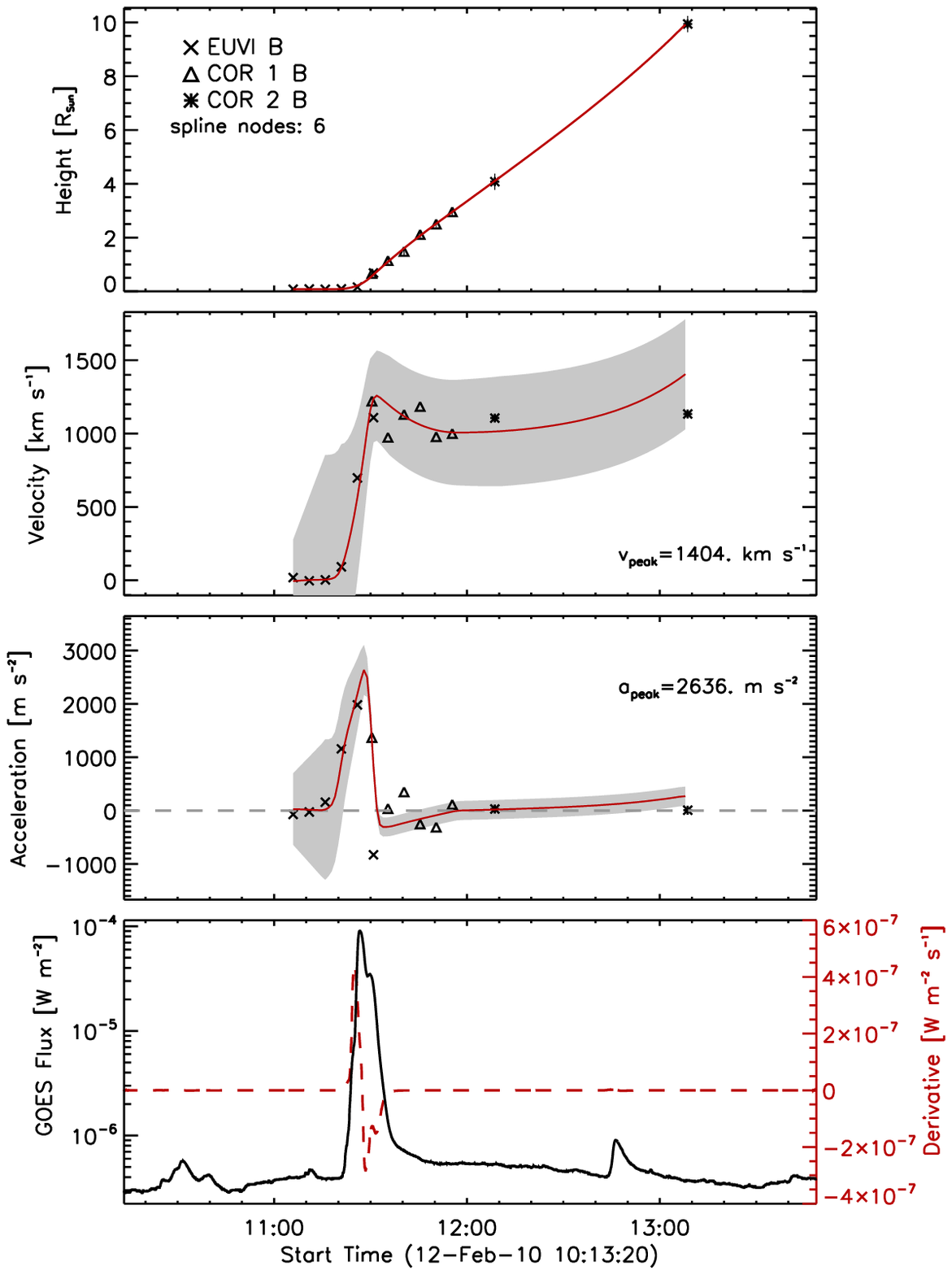}
	\caption{Same as Figure \ref{height1} but for the events observed on 2009 December 22 (left) and 2010 February 12 (right).}
	\label{height2}
\end{figure*}

We studied 95 CMEs, which occurred between January 2007 and May 2010, i.e.\ at the transition from solar cycle no. 23 to no.\ 24. For each event the full CME height, velocity and acceleration profiles could be derived. For details on the methods applied we refer to paper I. For each event, which was associated with a GOES flare, we compared the CME kinematics with the GOES SXR flux evolution and its derivative. Because of the temporal correlation between the HXR emission caused by flare-accelerated electron beams and the derivative of the SXR emission observed in many flare events \citep[Neupert effect; ][]{neupert1968, dennis1993, veronig2002, veronig2005} we use the derivative of the GOES 1--8~\AA~flux as a proxy for the evolution of the flare energy release. Figures \ref{height1} and \ref{height2} show four representative events of our sample. The top panels show the CME height, the second panels the CME velocity and the third panels the CME acceleration curve against time. In the bottom panels the GOES 1--8~\AA~flux and its derivative is plotted. 

For each CME we derived the following characteristic parameters: (1) peak velocity $v_{max}$, (2) peak acceleration $a_{max}$, (3) acceleration phase duration $t_{acc}$, (4) first measured height $h_{0}$, (5) height at peak velocity $h_{vmax}$, (6) height at peak acceleration $h_{amax}$. We applied a spline fit to the CME height-time profile and from its first and second derivative we derived $v_{max}$ and $a_{max}$, respectively. $a_{max}$ is given by the maximum value of the CME acceleration profile; $v_{max}$ is defined as the value of the velocity profile where the acceleration profile has decreased to 10\% of its peak value. This definition is used in order to ensure that the peak velocity during the CME impulsive acceleration phase \citep{zhang2004} is calculated excluding the effect of the subsequent residual acceleration phase (for details see paper I). The acceleration phase duration $t_{acc}$ was defined as the time difference between $t_{acc\_end}$, the time when the acceleration profile has decreased to 10\% of its maximum value, and $t_{acc\_start}$, the time in the increasing phase of the acceleration when the profile has reached 10\% of its maximum value. $h_{0}$, the first measured height in our height-time plots, is used as a rough estimate of the CME intiation height, because we measured the CMEs as soon as it was possible to observe them in the low corona. The heights $h_{vmax}$ and $h_{amax}$ are defined as the heights where the CME reached its peak velocity ($v_{max}$) and peak acceleration ($a_{max}$). A detailed study of these CME parameters is presented in paper I.

\subsection{CME characteristics in dependence of flare/filament association}

\begin{figure}
	\centering
		\includegraphics[scale=1.1]{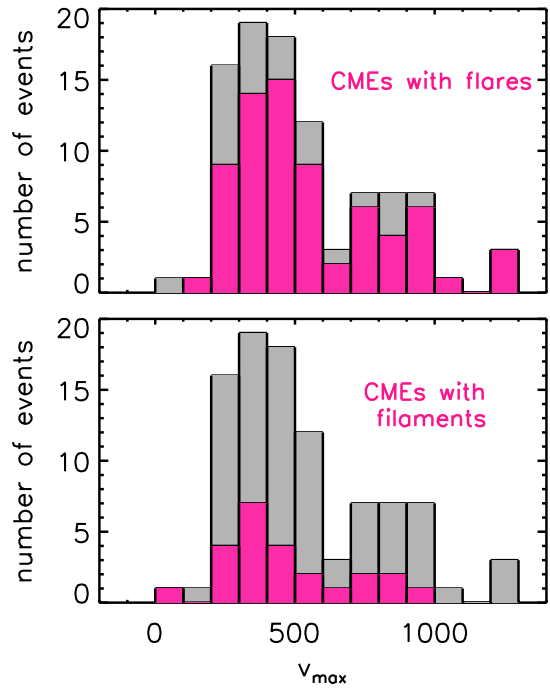}
	\caption{Histogram of the CME peak velocity $v_{max}$ for the whole sample of 95 events (grey distributions in the top and bottom panels). In the top panel the $v_{max}$ distribution of CME events associated with a flare is overlaid in color, in the bottom panel the histogram of CME events associated with an erupting filament is overlaid.}
	\label{histvmax}
\end{figure}

\begin{figure}
	\centering
		\includegraphics[scale=1.1]{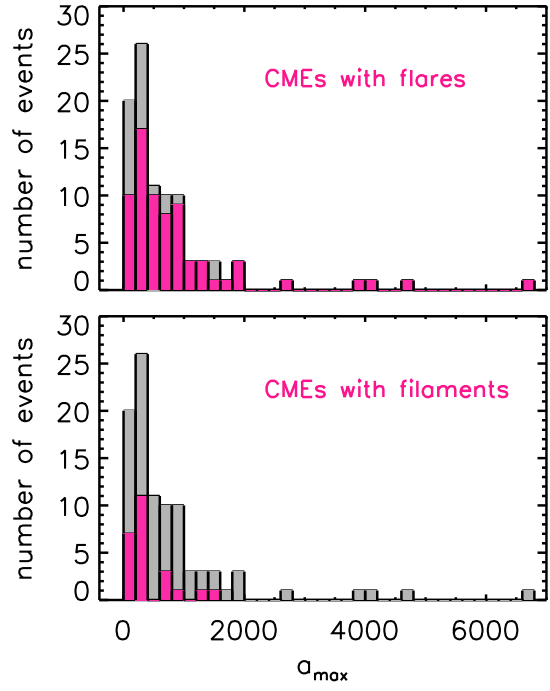}
	\caption{Same as Fig. \ref{histvmax} but for the CME peak acceleration.}
	\label{histamax}
\end{figure}

\begin{figure}
	\centering
		\includegraphics[scale=1.1]{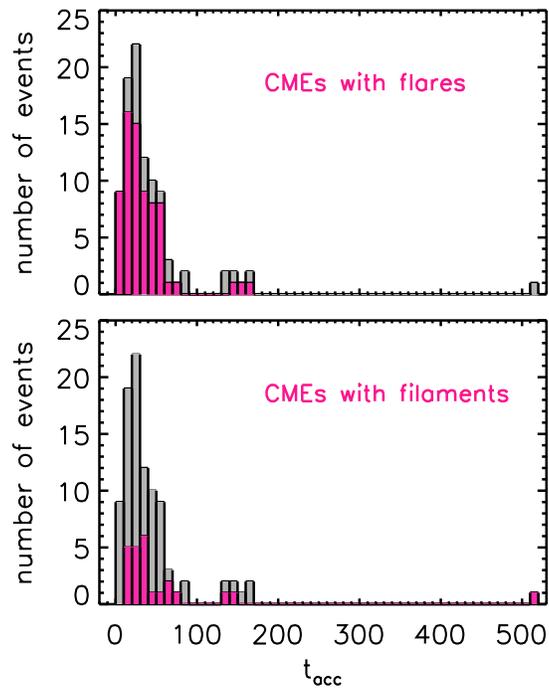}
	\caption{Same as Fig. \ref{histvmax} but for the CME acceleration duration.}
	\label{histaccdur}
\end{figure}

\begin{figure}
	\centering
		\includegraphics[scale=1.1]{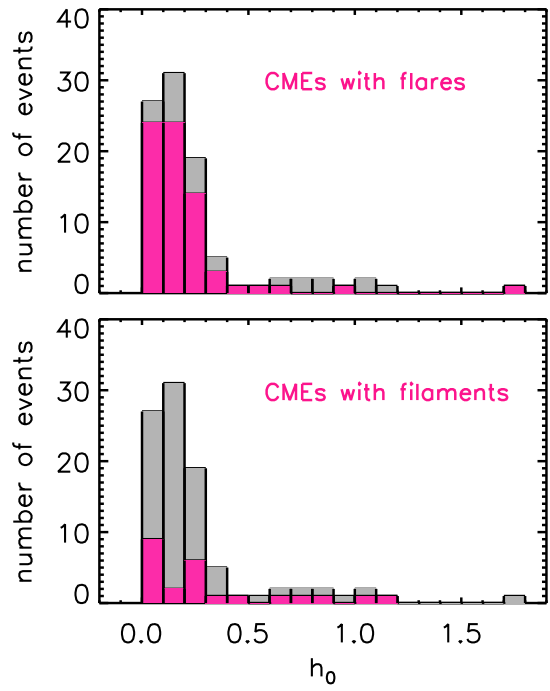}
	\caption{Same as Fig. \ref{histvmax} but for the height $h_{0}$ where the CME could be first identified. This height is used as an estimate of the CME initiation height.}
	\label{histh0}
\end{figure}

\begin{figure}
	\centering
		\includegraphics[scale=1.1]{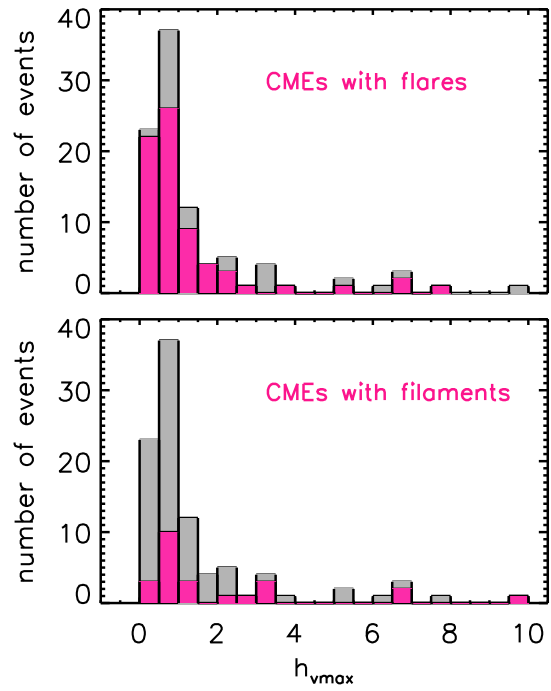}
	\caption{Same as Fig. \ref{histvmax} but for the height $h_{vmax}$, where the CMEs reached their maximum velocity. }
	\label{histhv}
\end{figure}

\begin{figure}
	\centering
		\includegraphics[scale=1.1]{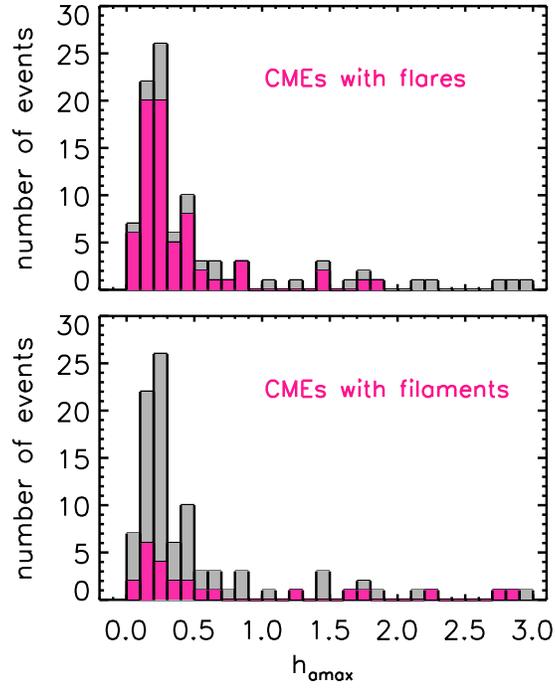}
	\caption{Same as Fig. \ref{histvmax} but for the height $h_{amax}$, where the CMEs reached their maximum acceleration. }
	\label{histha}
\end{figure}

We studied each CME parameter ($v_{max}$, $a_{max}$, $t_{acc}$, $h_{0}$, $h_{vmax}$ and $h_{amax}$) separately for CMEs with/without associated flares and with/without associated filament eruptions. Figures \ref{histvmax}--\ref{histha} show the distributions of the different CME parameters for the whole sample of 95 events (grey histograms) together with the distribution for events associated with flares (colored histogram in the top panel) and for events associated with filament eruptions (colored histogram in the bottom panel). For each parameter we derived the arithmetic mean together with its standard deviation and the median and its absolute deviation, for the whole sample as well as separately for the samples with/without associated flares or filaments. The resulting values are summarized in Table \ref{tab:tablesubgroups}. Because the distributions do not show a Gaussian behaviour but have a tail towards high values, a lognormal distribution better describes their behaviour \citep{limbert2001, yur2005, bein2011}. Thus we also calculated the mean value $\mu$ and the standard deviation $\sigma$ of the natural logarithm of the different CME parameter distributions \citep{cowan1998}:
\begin{equation}
\sigma=\sqrt{ln\left(\frac{Var}{E^2}+1\right)}
\end{equation}
\begin{equation}
\mu=ln(E)-\frac{\sigma^2}{2}
\end{equation}
with E the mean value and Var the variance of the quantity. In the following we use $\mu^*=e^\mu$ and $\sigma^*=e^\sigma$ as the median and multiplicative standard deviation of the lognormal probability function of the distribution \citep{limbert2001}. The confidence interval of 68.3\% is given by $e^{\mu\mp\sigma}$; $\mu$ and $\sigma$ are also summarized for each CME parameter in Table \ref{tab:tablesubgroups}.

To check if the CME distributions with/without associated flares and with/without associated filament eruptions, respectively, come from the same continuous distribution (null hypothesis) we used the Kolmogorov-Smirnov test \citep[e.g.][]{young1977, sachs1997}. This test is nonparametric,  i.e.\ it is not necessarily required to know the type of distribution (e.g.\ gaussian, lognormal, exponential). A test statistic D is calculated by
\begin{equation}
D=max\left|\frac{F_1}{n_1}-\frac{F_2}{n_2}\right|
\end{equation}
with $F_1$ and $F_2$ the cumulative distribution functions of both subsamples and $n_1$ and $n_2$ the number of events in each sample. D is then compared with
\begin{equation}
D_\alpha=K_\alpha\sqrt{\frac{n_1+n_2}{n_1 \cdot n_2}}.
\end{equation}
where K$_\alpha$ can be found in tables for different significance levels $\alpha$. $\alpha$ represents the probability that we wrongly reject the true hypothesis or accept the false hypothesis \citep{essenwanger1976}. If D$>$D$_\alpha$ the null hypothesis can be rejected, i.e.\ both subsamples do \textit{not} come from the same population.

\subsubsection{CME - Flare Relation}
\begin{figure}
	\centering
		\includegraphics[scale=0.5]{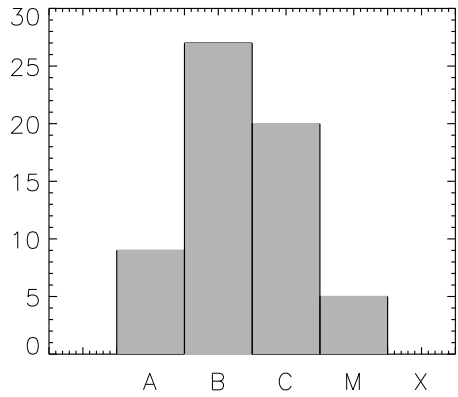}
	\caption{Distribution of the GOES classes of the flare events under study.}
	\label{goesclass}
\end{figure}

For 70 events out of our sample of 95 CMEs, we could identify an associated GOES class flare. Figure \ref{goesclass} shows the distribution of the GOES classes of the associated flares, which were not partially occulted (61 events): 9 events are associated with an A class flare, 27 with a B class flare, 20 events with a C class flare, and 5 events with an M class flare. None of the CME events under study is associated with an X class flare. 

In the upper panels of Figures \ref{histvmax}--\ref{histha} the distributions of CMEs associated with flares (70 events) are overplotted in color. On average, the CME peak velocity $v_{max}$ is higher for events associated with flares than for those with no flare association. Assuming a lognormal distribution we find $\mu^*=495^{-95}_{+117}$~km~s$^{-1}$ for events associated with flares and $\mu^*=406^{-89}_{+113}$~km~s$^{-1}$ for events with no flare-association. All CMEs, which reached a velocity higher than 1000 km~s$^{-1}$ are associated with a flare. These results are in accordance with former studies. \citet{burkepile2004}, who measured the average velocity of 111 limb CMEs observed with the SMM coronagraph/polarimeter, found mean values of 566 $\pm$ 67~km~s$^{-1}$ for CMEs with and 444 $\pm$ 59~km~s$^{-1}$ for CMEs without associated flares. \citet{vrsnak2005} analysed the CME mean velocity and the velocity at a distance of 3 $R_{\odot}$ of CMEs using LASCO data and found for the flare associated CME sample higher mean values than for the whole sample of CMEs. However, the Kolmogorov-Smirnov test applied to our sample does not show a clear distinction of both subsamples up to a significance level of 0.2.

For the CME peak acceleration $a_{max}$ the differences in mean values and median values for events associated with flares and events with no flare association are even higher than for the velocities (cf.\ Table \ref{tab:tablesubgroups}). We find a more than twice as high $\mu^*=551^{-343}_{+904}~$m~s$^{-2}$ for events with flare association than for CMEs without a flare, $\mu^*=240^{-137}_{+321}~$m~s$^{-2}$. Although the range of $a_{max}$ values for flare associated CMEs is large (77--6781~m~s$^{-2}$), all events which reached a peak acceleration higher than 1600 m~s$^{-2}$ (9 events out of our sample of 95 CMEs) were associated with a flare. The Kolmogorov-Smirnov test shows for the parameter $a_{max}$ at a significance level of 0.05 that both distributions (with/without associated flare) do not come from the same population.

For the acceleration phase duration, we found $\mu^*=25.3^{-11.2}_{+20.0}$ min for events with associated flares, compared to the $\mu^*=43.8^{-28.6}_{+82.4}$ min for events with no associated flare. All CMEs with $t_{acc}\leq$14~min were associated with a flare. Applying the Kolmogorov-Smirnov test it can be shown that both distributions do not come from the same population at a 0.10 significance level.

The measured CME height parameters $h_{0}$, $h_{vmax}$ and $h_{amax}$ are on average smaller for CMEs associated with flares than for CMEs with no flare (see Table \ref{tab:tablesubgroups}). $\mu^*$ of $h_{0}$ and $h_{vmax}$ are more than twice as high for events with no flare than for events with an associated flare. For $\mu^*$ of $h_{amax}$ the differences between these subgroups are lower (ratio $\sim$ 1.32). The Kolmogorov-Smirnov test showed for all height parameters that both subgroups do not come from the same population at a significance level of 0.10 for $h_0$, 0.15 for $h_{vmax}$ and even 0.05 for $h_{amax}$.

\subsubsection{CME - Filament Relation}

24 out of 95 CMEs were associated with an erupting filament. In the bottom panels of Figures \ref{histvmax}--\ref{histha} the distributions of these events are overplotted in color. Events, which could be associated with filaments showed lower $v_{max}$ than events without filament eruptions. For the former group we find $\mu^*=420^{-67}_{+78}$~km~s$^{-1}$, whereas for the latter group of CMEs we obtain $\mu^*$ of $v_{max}$ of 531$^{-107}_{+134}$~km~s$^{-1}$. \citet{burkepile2004} found no significant difference between both subgroups for the mean velocity of $\sim$520~km~s$^{-1}$.

Similar to the CME-flare association, the differences in the mean and median values for $a_{max}$ are more significant than for $v_{max}$. CMEs which could be associated with filaments showed lower $a_{max}$, with $\mu^*$ of 281$^{-136}_{+264}$ m~s$^{-2}$. None of these events reached $a_{max}$ higher than 1600 m~s$^{-2}$, 75\% have $a_{max}$ $<$400 m~s$^{-2}$. CME events without associated filament eruption show $\mu^*$ of 543$^{-337}_{+915}$ m~s$^{-2}$, i.e. the difference between these two subgroups is about a factor of 2. $a_{max}$ was the only parameter, for which the  Kolmogorov-Smirnov test suggested that both distributions (filament/non-filament associated events) do not come from the same population at a significance level of 0.05.

For the CME acceleration duration $t_{acc}$ we found $\mu^*$ of 33.1$^{-24.00}_{+87.00}$ min for events with associated and $\mu^*$ of $27.8^{-13.10}_{+24.74}$ min for events with no filament association. Because filament eruptions are correlated to CMEs with larger $a_{max}$ it is not surprising that they are also correlated to events with longer $t_{acc}$, due to the strong negative correlation between $a_{max}$ and $t_{acc}$ \citep[see][]{vrsnak2007, bein2011}. 

All CME height parameters $h_{0}$, $h_{vmax}$ and $h_{amax}$ have on average larger values for CMEs associated with filament eruptions than for CMEs without filament eruptions (Table \ref{tab:tablesubgroups}). The differences in the median values of the lognormal distribution are about a factor of 1.5.

~\newline
Considering the mean and median values of CME parameters we found higher values for $v_{max}$ and $a_{max}$ and smaller values for $t_{acc}$, $h_{0}$, $h_{vmax}$ and $h_{amax}$ for events with associated flares than for those with no flare. This effect is smallest for the CME peak velocity. The differences in the mean and median values show that CMEs associated with flares tend to have a more impulsive and intense acceleration and start from lower heights in the corona. CMEs which could be associated with an erupting filament behave oppositely. Their mean and median values of $v_{max}$ and $a_{max}$ are lower than for CMEs with no erupting filament and the mean and median values of $t_{acc}$, $h_{0}$, $h_{vmax}$ and $h_{amax}$ are higher, when a filament eruption was associated. The Kolmogorov-Smirnov test shows for all CME parameters but for $v_{max}$ that the distributions of flare/non-flare associated events do not come from the same population. Considering filament/non-filament events the Kolmogorov-Smirnov test shows only for $a_{max}$ a distinctions between both distributions. 

\begin{table*}
	\centering
	\scriptsize
		\begin{tabular}{||c|c|c|c|c|c|c||}
		\hline
		\hline
			~&&all & with & without & with & without \\
			~&& CMEs& GOES flare & GOES flare& filament & filament \\
				~&& (95 events)& (70 events) & (25 events) & (24 events) & (71 events) \\
			\hline
			
			$v_{max}$ [km~s$^{-1}]$ & mean $\pm$ s & 526 $\pm$ 263 & 550 $\pm$ 268 & 459 $\pm$ 242 & 467 $\pm$ 227 & 546 $\pm$ 273 \\
			& median $\pm$ mad &460 $\pm$ 160& 461 $\pm$ 150 & 393 $\pm$ 135 & 400 $\pm$ 121 &  461 $\pm$ 161\\
			& $\mu\pm\sigma$ & 6.15 $\pm$ 0.22 & 6.20 $\pm$ 0.12 & 6.00 $\pm$ 0.25 & 6.04 $\pm$ 0.21 & 6.19 $\pm$ 0.22\\
			\hline
			$a_{max}$ [m~s$^{-2}]$ & mean $\pm$ s & 757 $\pm$ 1034& 896 $\pm$ 1148 & 367 $\pm$ 424 & 391 $\pm$ 380 &  880 $\pm$ 1152 \\
			& median $\pm$ mad & 414 $\pm$ 246 & 556 $\pm$ 306 & 241 $\pm$ 134 & 266 $\pm$ 103 & 545 $\pm$ 321 \\
				& $\mu\pm\sigma$ & 6.10 $\pm$ 1.05 & 6.31 $\pm$ 0.97 & 5.48 $\pm$ 0.85 & 5.64 $\pm$ 0.66 & 6.28 $\pm$ 1.00\\
			\hline
			$t_{acc}$ [min]  & mean $\pm$ s & 44.6 $\pm$ 60.4 & 33.9 $\pm$ 30.2 & 74.4 $\pm$ 102.0 & 63.1 $\pm$ 102.3 & 38.3 $\pm$ 36.1 \\
			& median $\pm$ mad & 29.0 $\pm$ 14.5 & 26.0 $\pm$ 13.5 & 33.0 $\pm$ 14.5 & 35.0 $\pm$ 17.0 & 26.0 $\pm$ 16.0 \\
				& $\mu\pm\sigma$ & 3.28 $\pm$ 1.04 & 3.23 $\pm$ 0.58 & 3.77 $\pm$ 1.06 & 3.50 $\pm$ 1.29 & 3.33 $\pm$ 0.64\\
			\hline
			$h_{0}$ [$R_{\odot}$] & mean $\pm$ s & 0.24 $\pm$  0.29 & 0.19 $\pm$ 0.24 & 0.40 $\pm$ 0.35 & 0.31 $\pm$ 0.33 & 0.22 $\pm$ 0.28 \\
			& median $\pm$ mad & 0.14 $\pm$  0.08 & 0.12 $\pm$ 0.06 & 0.27 $\pm$ 0.16 & 0.23 $\pm$ 0.17 & 0.14 $\pm$ 0.07 \\
			& $\mu\pm\sigma$ & $-$1.85 $\pm$ 0.88 & $-$2.17 $\pm$ 0.99 & $-$1.19 $\pm$ 0.57 & $-$1.56 $\pm$ 0.77 & $-$1.97 $\pm$ 0.93\\
			\hline
			$h_{amax}$ [$R_{\odot}$] & mean $\pm$ s & 0.53 $\pm$  0.64 & 0.36 $\pm$ 0.37 & 0.99 $\pm$ 0.95 & 0.72 $\pm$ 0.87 & 0.46 $\pm$ 0.53 \\
			& median $\pm$ mad & 0.26 $\pm$  0.12 & 0.24 $\pm$ 0.10 & 0.52 $\pm$ 0.34 & 0.35 $\pm$ 0.21 & 0.26 $\pm$ 0.12 \\
			& $\mu\pm\sigma$ & $-$1.10 $\pm$ 0.90 & $-$1.38 $\pm$ 0.72 & $-$0.34 $\pm$ 0.65 & $-$0.79 $\pm$ 0.90 & $-$1.20 $\pm$ 0.84\\
		 \hline
		 $h_{vmax}$ [$R_{\odot}$] & mean $\pm$ s & 1.56 $\pm$  1.82 & 1.17 $\pm$ 1.48 & 2.26 $\pm$ 2.38 & 2.00 $\pm$ 2.42 & 1.28 $\pm$ 1.54 \\
			& median $\pm$ mad & 0.78 $\pm$ 0.42 & 0.70 $\pm$ 0.35 & 1.05 $\pm$ 0.53 & 0.86 $\pm$ 0.40 & 0.76 $\pm$ 0.41 \\
			& $\mu\pm\sigma$ & $-$0.09 $\pm$ 0.94 & $-$0.32 $\pm$ 0.96 & 0.44 $\pm$ 0.75 & 0.24 $\pm$ 0.91 & $-$0.21 $\pm$ 0.90\\
			\hline
		 \hline
		\end{tabular}
		\caption{Arithmetic mean with standard deviation, median with median absolute deviation (mad) and mean value $\mu$ with standard deviation $\sigma$ of the lognormal probability function of the following decisive CME parameters: peak velocity $v_{max}$, peak acceleration $a_{max}$, acceleration phase duration $t_{acc}$, height $h_{0}$ where the CME leading edge could be identified for the first time, height at peak acceleration $h_{amax}$, height at peak velocity $h_{vmax}$.  Mean, median and $\mu$ of all these parameters were derived for the whole sample of events (third column) and for several subgroups: events with associated flares (fourth column), events for which no flare could be associated (fifth column), events with associated filament eruptions (sixth column), events for which no filament eruptions could be associated (seventh column).\newline}
\label{tab:tablesubgroups}		
\end{table*}

\subsection{Relationship between CME and flare parameters}

\begin{figure}
	\centering
		\includegraphics[scale=1]{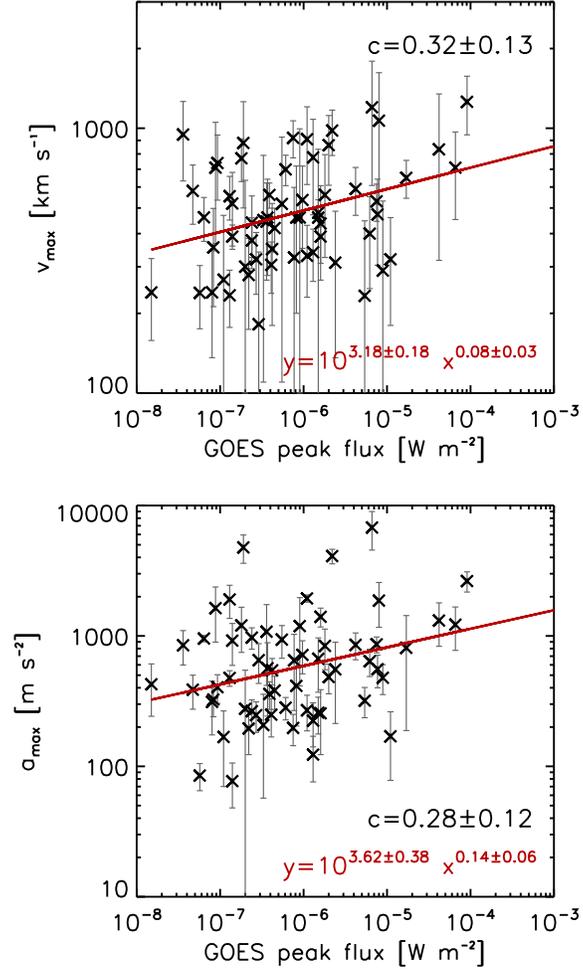}
	\caption{CME peak velocity (top) and peak acceleration (bottom) against the GOES peak flux of the associated flare. A double logarithmic space is used for the plot and the calculation of the correlation coefficient $c$, which is annotated in each panel. The regression line is overplotted in red.}
	\label{goesmax}
\end{figure}

We analysed the scaling of the GOES peak flux $F_{SXR}$ and the GOES SXR rise time $t_{SXR}$ with the characteristics of the associated CMEs, i.e. $v_{max}$, $a_{max}$, $t_{acc}$, $h_{0}$, $h_{vmax}$ and $h_{amax}$ and calculated their dependency in logarithmic space. For these correlations we only considered those 61 (out of 70) flare events, which were not partially occulted from Earth view. The correlations are in general weak. Figures \ref{goesmax} and \ref{risetime} show four representative correlations.

We used a bootstrap method \citep{efron1979} within a Monte Carlo algorithm to estimate the standard error of our calculated correlation coefficients \citep{efron1986}. A bootstrap sample is constructed by randomly drawing with replacement from the original data pairs, with the bootstrap sample having the same number of entries than the original sample. Such bootstrap samples are constructed 1000 times and for each bootstrap sample the correlation coefficient is calculated. The mean value plus standard deviation of all these 1000 calculations is used as a robust measure of the 
correlation coefficient and its standard error.

\begin{figure}
	\centering
		\includegraphics[scale=0.7]{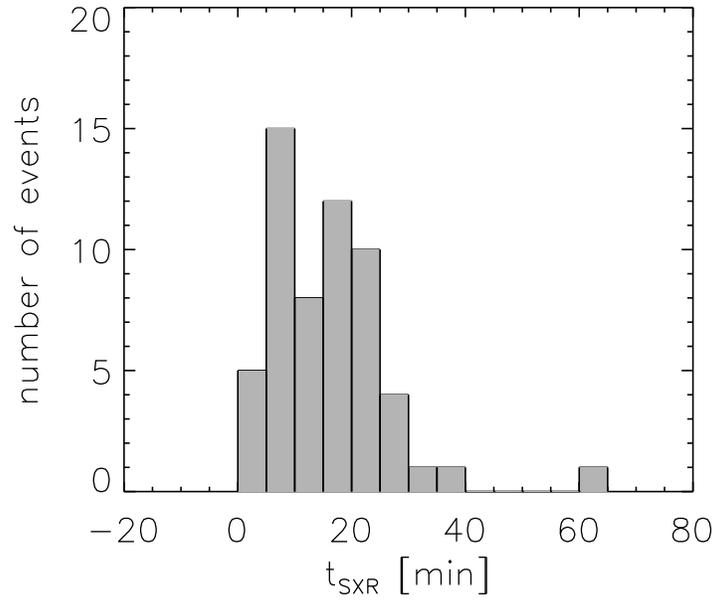}
	\caption{Distribution of the risetime $t_{SXR}$ of the GOES SXR flares.}
	\label{histrisetime}
\end{figure}

We found a positive correlation between $F_{SXR}$ and $v_{max}$ of $c=$ 0.32$\pm$0.13 (bottom panel of Figure \ref{goesmax}), i.e.\ CMEs which were associated with flares with higher SXR flux tend to reach higher peak velocities. This is consistent with the results from \citet{maricic2007} who analysed the correlation between SXR flares and CMEs for a set of 18 events, including B to X class flares, finding $c=$0.52. 
The power law dependency between $F_{SXR}$ and $v_{max}$ of our study can be expressed by:
\begin{equation}
	v_{max}=10^{3.18\pm0.18}~F_{SXR}^{0.08\pm0.03}.
\end{equation}
The correlation between $F_{SXR}$ and $a_{max}$ is $c=$0.28$\pm$0.12, see the top panel of Figure \ref{goesmax}, i.e.\ CMEs with high peak accelerations tend to be associated with flares of high GOES SXR flux. \citet{maricic2007} also found a positve correlation between these two parameters with a correlation coefficient of $c$=0.60. We found a power law dependency of:
\begin{equation}
	a_{max}=10^{3.62\pm0.38}~F_{SXR}^{0.14\pm0.06}.
\end{equation}
We determined the SXR rise time $t_{SXR}$ from the derivative of the GOES flux curves as the time difference between the start of the increase and the maximum of the SXR derivative, which was possible for 57 events. The distribution of the time differences is shown in Figure \ref{histrisetime}. The GOES SXR rise time $t_{SXR}$ and CME parameters (Figure \ref{risetime}) do not yield distinct correlations. The highest correlation coefficient was obtained for the relation between $t_{SXR}$ and the CME acceleration duration, $c=$ 0.37$\pm$0.15. The power low dependency between these two parameters can be expressed as:
\begin{equation}
	t_{acc}=10^{1.98\pm0.39}~t_{SXR}^{0.40\pm0.13}.
\end{equation}
Between $a_{max}$ and $t_{SXR}$ we found a negative correlation with $c=-0.32\pm0.15$ and a power law dependency of
\begin{equation}
	a_{max}=10^{4.02\pm0.48}~t_{SXR}^{-0.43\pm0.16},
\end{equation}
i.e.\ CMEs with high peak accelerations are preferentially associated with flares, which show short SXR rise times.

One possible explanation for the weak flare-CME correlations we obtained is the lack of strong flare events in our sample which contains only a few M class flares and no X class flares due to the solar activity minimum condition in the period under study (2007--2010). Since we cover only low energetic flares and thus a relatively narrow range of GOES classes, possible correlations may be hidden in the scatter of the data points. 
It turned out that weak flares show a larger scatter in the distribution of values which weakens our correlation results since predominantly A and B class flares are studied in our sample. To circumvent this effect we calculated the mean and median value of the CME parameters separately for each GOES class (shown in Table \ref{tab:tablegoesclass}). By this procedure the CME parameters were weighted for each GOES class equally which reduced the influence of the scattering of the large number of A- and B-class flares in our sample.

\begin{figure}
	\centering
		\includegraphics[scale=1]{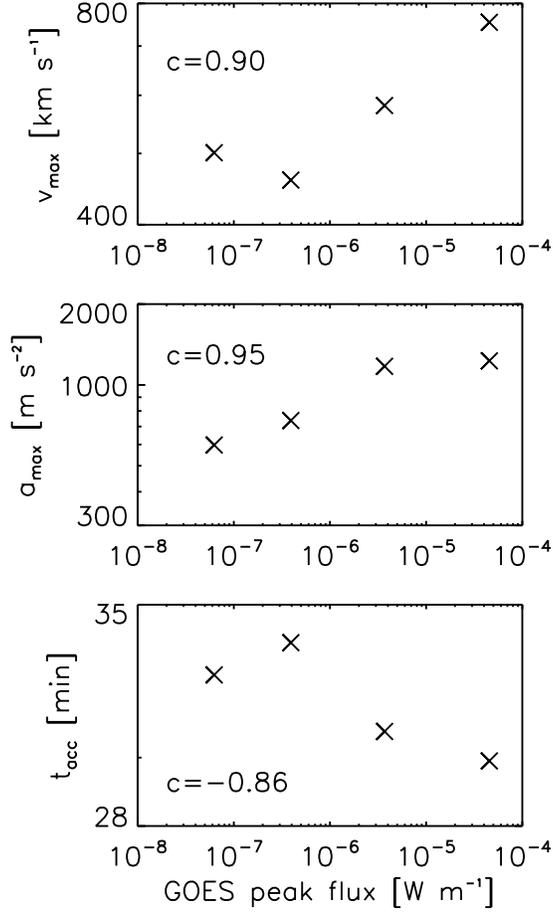}
	\caption{Mean value of the CME peak velocity $v_{max}$ (top), acceleration $a_{max}$ (middle) and acceleration duration $t_{acc}$ (bottom) for each GOES class plotted against the mean GOES flux in each subgroup. The correlation coefficient $c$ (calculated in logarithmic space) is annotated in each panel.}
	\label{goessep}
\end{figure}

Figure \ref{goessep} shows these mean values for $v_{max}$, $a_{max}$ and $t_{acc}$ plotted against the mean GOES flux of each subgroup. In this representation a clear trend is noticeable that CMEs, which reach higher peak velocities are related to flares with higher GOES peak flux (top panel in Figure \ref{goessep}), consistent with former studies. \citet{moon2002} analysed 3217 LASCO CMEs, which were associated with flares and compared the median value of the CME speeds derived from the whole sample, from CMEs associated with flares $>$C1 and from CMEs associated with flares $>$M1. When considering the whole sample the authors found the lowest median value, whereas CMEs, which were associated with flares $>$M1, showed the largest median value for the CME speed distribution. \citet{vrsnak2005} compared the velocities between B and C flare associated CMEs with the velocities of M- and X-class flare associated CMEs and found that the mean value of the M- and X-class flare associated events is about 1.4 times of the mean value of the other subgroup. 

We also find that the mean values of the CME peak acceleration is positively correlated with the GOES peak flux (middle panel of Figure \ref{goessep}). The bottom panel of Figure \ref{goessep} shows the mean values of the CME acceleration duration against the GOES peak flux, which shows a tendency that $t_{acc}$ is smaller for CMEs associated with higher energetic flares. The mean and median values calculated for each GOES class separately for all CME parameters under study are summarized in Table \ref{tab:tablegoesclass}. 

\begin{table*}
\centering
\begin{tabular}{||c|c|c|c|c|c||}
		\hline
		\hline
		 &~&A-class & B-class & C-class & M-class\\
		 &~& (9 events) & (27 events) & (20 events) & (5 events)\\
		 \hline
		 $v_{\rm{max}}$& mean $\pm$ s& 501$\pm$258 & 532$\pm$257 & 609$\pm$301 & 754$\pm$340\\
		 ~& median $\pm$ mad& 461$\pm$221 & 460$\pm$135 & 530$\pm$200 & 710$\pm$122\\
		 $a_{\rm{max}}$&mean $\pm$ s& 597$\pm$474 & 986$\pm$1236 & 1229$\pm$1489 & 1230$\pm$906\\
		 ~& median $\pm$ mad & 404$\pm$92 & 639$\pm$370 & 810$\pm$502 & 1221$\pm$411\\
		 $t_{\rm{acc}}$&mean $\pm$ s& 32.6$\pm$15.6 & 28.9$\pm$24.1 & 25.2$\pm$15.7 & 29.9$\pm$17.7\\
		 ~& median $\pm$ mad & 26.0$\pm$13.0 & 21.0$\pm$12.0 & 20.5$\pm$9.5 & 20.5$\pm$7.5 \\
		 $h_{\rm{0}}$&mean $\pm$ s & 0.22$\pm$0.19 & 0.19$\pm$0.27 & 0.14$\pm$0.19 & 0.27$\pm$0.39\\
		 ~& median $\pm$ mad & 0.19$\pm$0.09 & 0.11$\pm$0.05 & 0.10$\pm$0.05 & 0.11$\pm$0.03 \\		
		 $h_{\rm{vmax}}$&mean $\pm$ s & 1.03$\pm$0.62 & 0.89$\pm$1.01 & 0.70$\pm$0.47 & 0.80$\pm$0.65\\
		 ~& median $\pm$ mad & 1.30$\pm$0.61 & 0.70$\pm$0.28 & 0.61$\pm$0.21 & 0.70$\pm$0.20 \\	
		 $h_{\rm{amax}}$&mean $\pm$ s & 0.41$\pm$0.28 & 0.31$\pm$0.33 & 0.31$\pm$0.32 & 0.58$\pm$0.58\\
		 ~& median $\pm$ mad & 0.33$\pm$0.17 & 0.23$\pm$0.08 & 0.22$\pm$0.10 & 0.27$\pm$0.13 \\	
		 \hline 
\end{tabular}
\caption{Mean value with standard deviation and median with the median absolute deviation (mad) of the CME parameters $v_{max}$, $a_{max}$, $t_{acc}$, $h_{0}$, $h_{vmax}$ and $h_{amax}$, separately for the flare GOES classes.}
\label{tab:tablegoesclass}
\end{table*}

\subsection{Flare-CME temporal relationship}

We studied the relative timing of the CME acceleration phase and the flare energy release in order to investigate possible temporal synchronizations between these two phenomena. From the CME observations we derived the start ($t_{acc\_start}$), peak ($t_{acc\_peak}$) and end time ($t_{acc\_end}$) of the CME acceleration. From the GOES flare observations we derived the start and peak time of the 1--8~\AA~SXR flux as well as the peak of the 1--8~\AA~flux derivative. For this study we used only flare events, which were not partially occulted and for which a peak in the derivative of the GOES flux could be derived. This was the case for 57 out of 70 flare-CME pairs. 

\begin{figure*}
	\centering
		\includegraphics[scale=0.75]{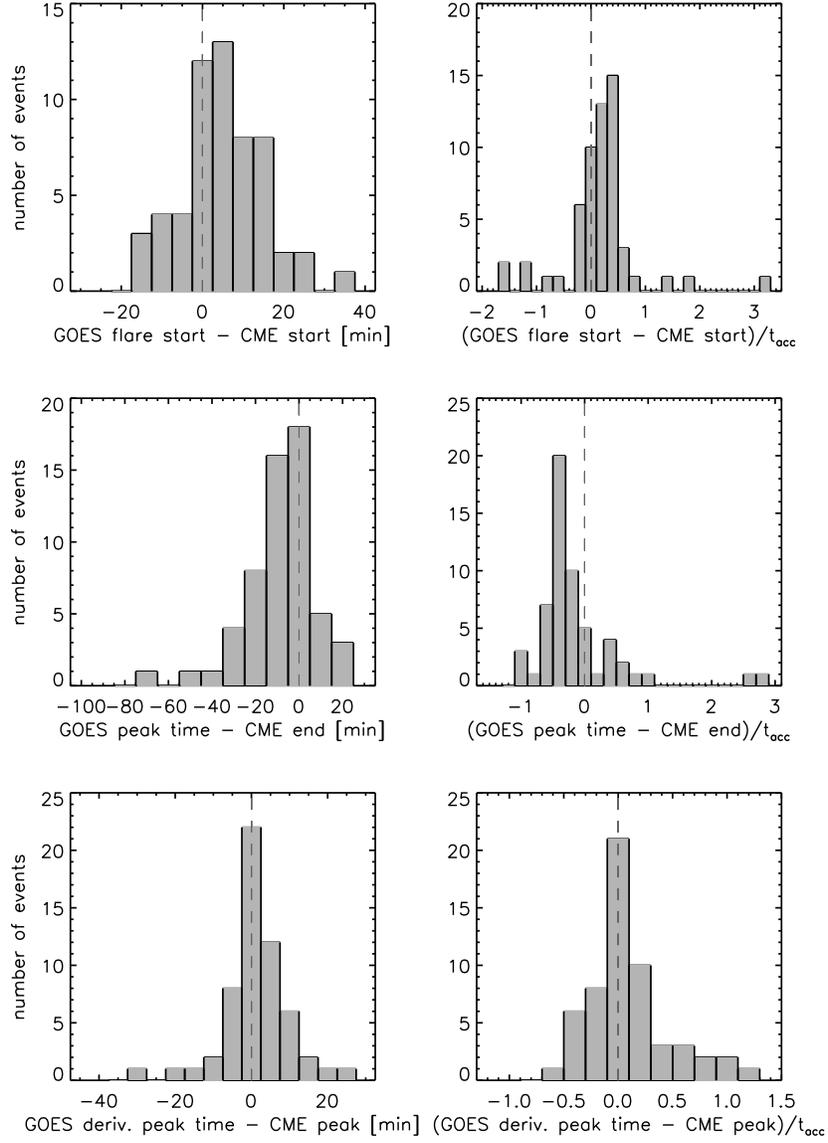}
	\caption{Distribution of the time differences between the start of the GOES flare and the start of the CME acceleration (top), GOES peak time and CME acceleration end time (middle) and the peak time of the derivative of the GOES 1--8 \AA~flux and the CME acceleration peak time (bottom). On the left hand side, the time differences are plotted in minutes, whereas on the right hand side the time differences are normalized by the acceleration duration of the corresponding CMEs. Positive values mean that the CME acceleration start, end or peak occurred before the flare SXR start, peak or derivative peak, respectively.}
	\label{histtime}
\end{figure*}

The top panels of Figure \ref{histtime} show the histograms of the time differences between the GOES flare start times and the start of the associated CME acceleration. For the distribution on the left hand side the differences expressed in minutes were used, whereas the values on the right hand side were normalized by the CME acceleration duration. For the time differences we found a mean value of 4.9 $\pm$ 10.2 min and a median of 4.6 $\pm$ 6.3 min. The distribution shows that the values are confined to within about $\pm$20 min (93\% of the events lie within this interval). In 75\% of the cases, the time difference is positive, i.e. the CME acceleration starts \textit{before} the rise of the flare soft X-ray flux. Considering the normalized values, the time difference is smaller than $\pm$0.5 for 77\% of the events and smaller than $\pm$0.25 for 44\%. \citet{maricic2007} also found that for the majority of the events the CME acceleration starts before the SXR flare with a mean value of 23 $\pm$ 30 min (for a set of 18 events containing a significant fraction of gradual events). This mean value is larger than in our study, but the time delays normalized by $t_{acc}$ are quite similar in both studies, with a mean value of 0.17 $\pm$ 0.71 in our study and 0.14 $\pm$ 0.20 in \citet{maricic2007}. 

In the middle panels of Figure \ref{histtime} we compare the end of the CME impulsive acceleration phase $t_{acc\_end}$ with the GOES SXR flare peak time, which relates to the end of the impulsive flare energy release phase. We found that for the majority of the events (77\%), the GOES peak occurred earlier than $t_{acc\_end}$. The mean value of the distribution is $-9.3$ $\pm$ 15.8 min and the median $-6.5$ $\pm$ $7.5$ min. For some gradual CMEs ($t_{acc}$ $>$50 min), the delay between flare peak and acceleration end can be $>$30 min (3 events). At the right hand side the normalized difference between the GOES peak flux and $t_{acc\_end}$ is plotted. The distribution shows an asymmetric behaviour with a tail at positive values,  and a distinct mode in the range of $-0.5$ to $-0.3$. For 65\% of the events, the normalized time differences are between $-0.7$ and $-0.1$.

The time difference between the peak of the GOES 1--8~\AA~flux derivative and the CME acceleration peak (bottom left panel of Figure \ref{histtime}) is found to be small. The distribution does strongly peak around zero; for the mean value we obtained 1.3 $\pm$ 8.6 min, for the median 1.0 $\pm$ 4.0 min. \citet{maricic2007}, who used the logarithmic derivative of the SXR GOES flux, found a mean value of 2.7 $\pm$ 14 min, consistent with our findings. In 81\% of the events, we found a difference smaller than $\pm$10~min, for 58\% of the events it is smaller than $\pm$5~min. Only one event of our study showed a time delay $>$20~min. 65\% of the normalized values (distribution shown at the right hand side of Figure \ref{histtime}) lie within $\pm$0.25. The statistical parameters for the time differences described above are summarized in Table \ref{tab:tabletimes}.

These findings provide strong evidence that the timing of the flare energy release and the CME dynamics (acceleration) are well synchronized, supporting previous results from case studies \citep[e.g.][]{zhang2004}. Recent studies also analysed the relationship between the flare HXR emission and the CME acceleration peak, finding that the flare HXR peak occurs close in time with the maximum CME acceleration \citep[][]{temmer2008, temmer2010, cheng2010b, berkebile2012}.

\begin{table*}
\centering
\footnotesize
\begin{tabular}{||c|c|c|c|c||}
		\hline
		\hline
		 Time Delay & Minimum & Maximum & Arithmetic Mean $\pm$ & Median \\
		  & &  &  Standard Deviation  & $\pm$ Mad\\
		 \hline
		 GOES flare start time $-$ $t_{acc\_start}$ [min]& $-17.3$ & 32.6 & 4.9 $\pm$ 10.2 & 4.6 $\pm$ 6.3 \\
		 (GOES flare start time $-$ $t_{acc\_start}$)/$t_{acc}$ & $-1.58$ & 3.26 & 0.17 $\pm$ 0.71 & 0.17$\pm$0.20 \\
		 GOES flare peak time $-$ $t_{acc\_end}$ [min] & $-71.5$ & 21.5 & $-9.3$ $\pm$ 15.8 & $-6.5$ $\pm$ 7.5\\
		 (GOES flare start time $-$ $t_{acc\_end}$)/$t_{acc}$ & $-1.08$ & 2.83 & $-0.13$ $\pm$ 0.69 & $-0.33$ $\pm$ 0.20 \\
		 GOES der. peak time $-$ $t_{acc\_peak}$ [min] & $-31.5$ & 23.0 & 1.29 $\pm$ 8.63 & 1.00 $\pm$ 4.00\\
		 (GOES der. peak time $-$ $t_{acc\_peak}$)/$t_{acc}$ & $-0.63$ & 1.17 & 0.09 $\pm$ 0.38 & 0.02 $\pm$ 0.18 \\
		 \hline 
		 \hline
\end{tabular}
\caption{Statistical parameters (minimum value, maximum value, arithmetic mean, standard deviation, median and median absolute deviation) for time delays between characteristic CME and flare times: time difference between the start of the GOES flare and the start of the CME acceleration $t_{acc\_start}$ (differences expresssed in minutes in the first row, values normalized by the CME acceleration time $t_{acc}$ in the second row), time delay between the GOES flare peak time and the end of the CME acceleration phase $t_{acc\_end}$ (delays expressed in minutes in the third row, normalized values in the forth row) and time difference between the peak time of the GOES derivative and the peak time of the CME acceleration $t_{acc\_peak}$ (delays expressed in minutes in the fifth, normalized values in the sixth row).}
\label{tab:tabletimes}
\end{table*}

\section{Summary and Conclusion}

Based on a sample of 95 CME events we present a statistical study on various characteristic CME parameters and their relation to flares. CMEs which are associated with flares show on average higher peak velocities ($v_{max}$), higher peak accelerations ($a_{max}$), shorter acceleration phase durations ($t_{acc}$), lower heights at peak velocity ($h_{vmax}$), lower heights at peak acceleration ($h_{amax}$) and lower initiation heights ($h_{0}$). The ratio between the median values of the lognormal probability functions of both subgroups is about a factor of 2. Only for $v_{max}$ the ratio is significantly smaller ($\sim$1.1), most probably due to the small range of $v_{max}$ values. Due to the anticorrelation between $a_{max}$ and $t_{acc}$ and the relation $v_{max}\approx a_{max}\cdot t_{acc}$ the range for $v_{max}$ is smaller than for the other two quantities, it basically covers only one order of magnitude. Although we found clear differences in the mean and median values of the two subgroups there exist also events associated with flares, which have low $v_{max}$ and $a_{max}$ values and high $t_{acc}$, $h_{vmax}$, $h_{amax}$ and $h_{0}$. For instance, the smallest measured $a_{max}$ value for a CME with flare was 77~m~s$^{-2}$ and the highest measured value for a CME without flare was 1577~m~s$^{-2}$. The Kolmogorov-Smirnov test suggested a distinction between both distributions (flare/non-flare associated events) for every CME parameter (except for $v_{max}$) at 0.05--0.15 levels of significance. The clearest distinctions were found for $a_{max}$ and $h_{amax}$ at a 0.05 level of significance.

CMEs which are associated with erupting filaments show on average smaller $v_{max}$, $a_{max}$, and larger $t_{acc}$, $h_{vmax}$, $h_{amax}$ and $h_{0}$. These trends are the other way round than for the flare association. The ratio between the mean and median values of both subgroups is somewhat smaller than for the flare association. Again we found the smallest ratio between the median values $\mu^*$ of $v_{max}$ (1.1). The $\mu^*$ of $a_{max}$ and $h_{vmax}$ showed the highest ratio with 1.9. For $t_{acc}$, $h_{amax}$ and $h_{0}$ we found ratios between 1.2 and 1.5. But there exist also CMEs associated with filament eruptions, which have high $v_{max}$ and $a_{max}$ values and low $t_{acc}$, $h_{vmax}$, $h_{amax}$ and $h_{0}$. For example the highest measured $a_{max}$ value for a CME which was associated with an erupting filament was 1561~m~s$^{-2}$, the smallest value for a CME event with no erupting filament was 35~m~s$^{-2}$ (the second lowest value of the whole distribution). Both of these events were not associated with a flare. The Kolmogorov-Smirnov test did not show a clear distinction between both subgroups. Only for $a_{max}$ the test suggested that both distributions do not come from the same population at a 0.05 level of significance.

The correlations obtained between $v_{max}$, $a_{max}$ and $t_{acc}$ with the GOES peak flux $F_{SXR}$ and the SXR rise time $t_{SXR}$ of the associated flare were low. We found a weak positive correlation between the CME acceleration duration $t_{acc}$ and the flare rise time $t_{SXR}$ ($c=0.37\pm0.15$) and a weak negative correlation between $a_{max}$ and $t_{SXR}$ ($c=-0.32\pm0.15$). Correlations between $v_{max}$ and $a_{max}$ with $F_{SXR}$ were $c=0.32\pm0.13$ and $c=0.28\pm0.12$ respectively. If the events are averaged and binned into the different GOES classes, the correlations are much more distinct. The mean values in each GOES class show an increasing trend for $v_{max}$ and $a_{max}$, and a decreasing trend for $t_{acc}$ with higher GOES flux. 

\begin{figure}
	\centering
		\includegraphics[scale=1]{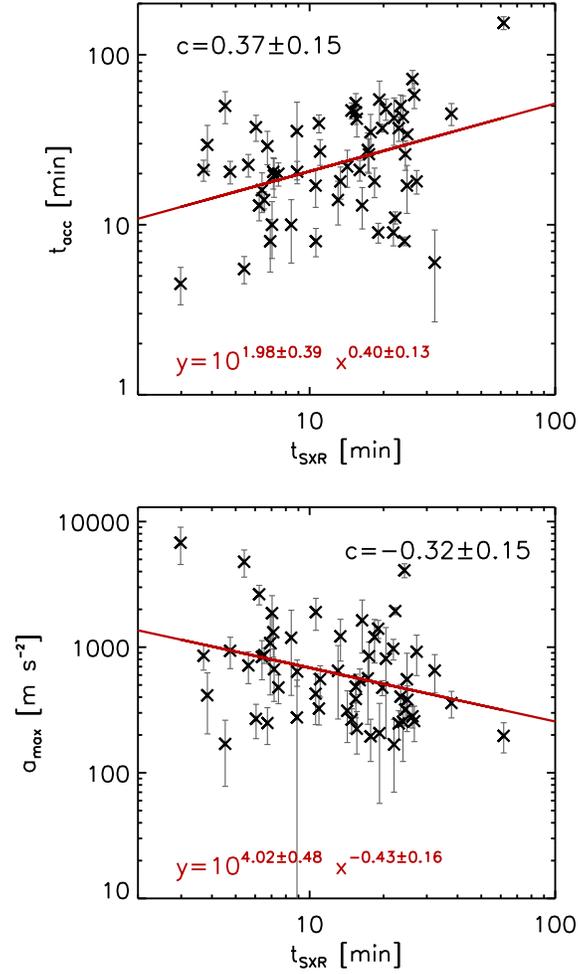}
	\caption{CME acceleration duration (top) and peak acceleration (bottom) against the flare SXR rise time. The correlation coefficient $c$ (calculated in double logarithmic space) and the regression line is overplotted. }
	\label{risetime}
\end{figure}

\begin{figure*}
	\centering
		\includegraphics[scale=0.75]{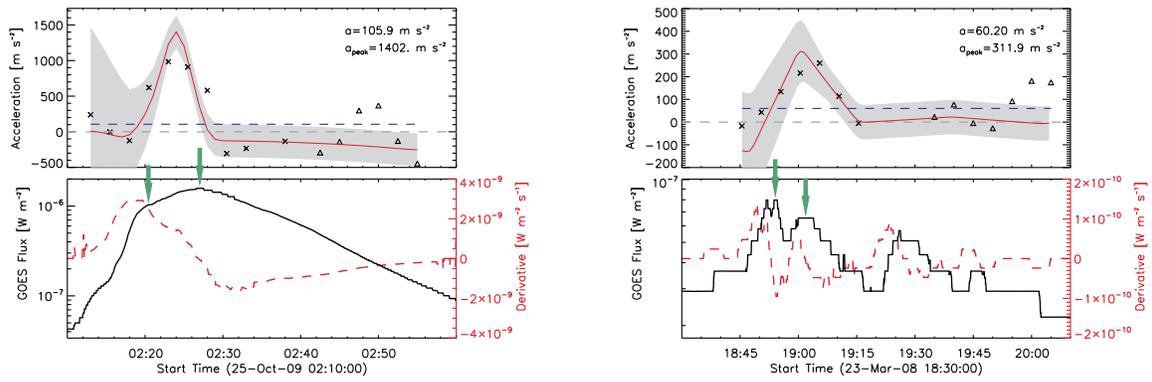}
	\caption{CME acceleration (top) and GOES SXR curve (bottom) from the events observed on 25 October 2009 (left hand side) and 23 March 2008 (right hand side). In both events the CME acceleration starts after the flare onset, which may be related to an overlap of two flares. Arrows mark two SXR maxima indicating two possible flare maxima.}
	\label{accsxr}
\end{figure*}

\begin{figure}
	\centering
		\includegraphics[scale=1.0]{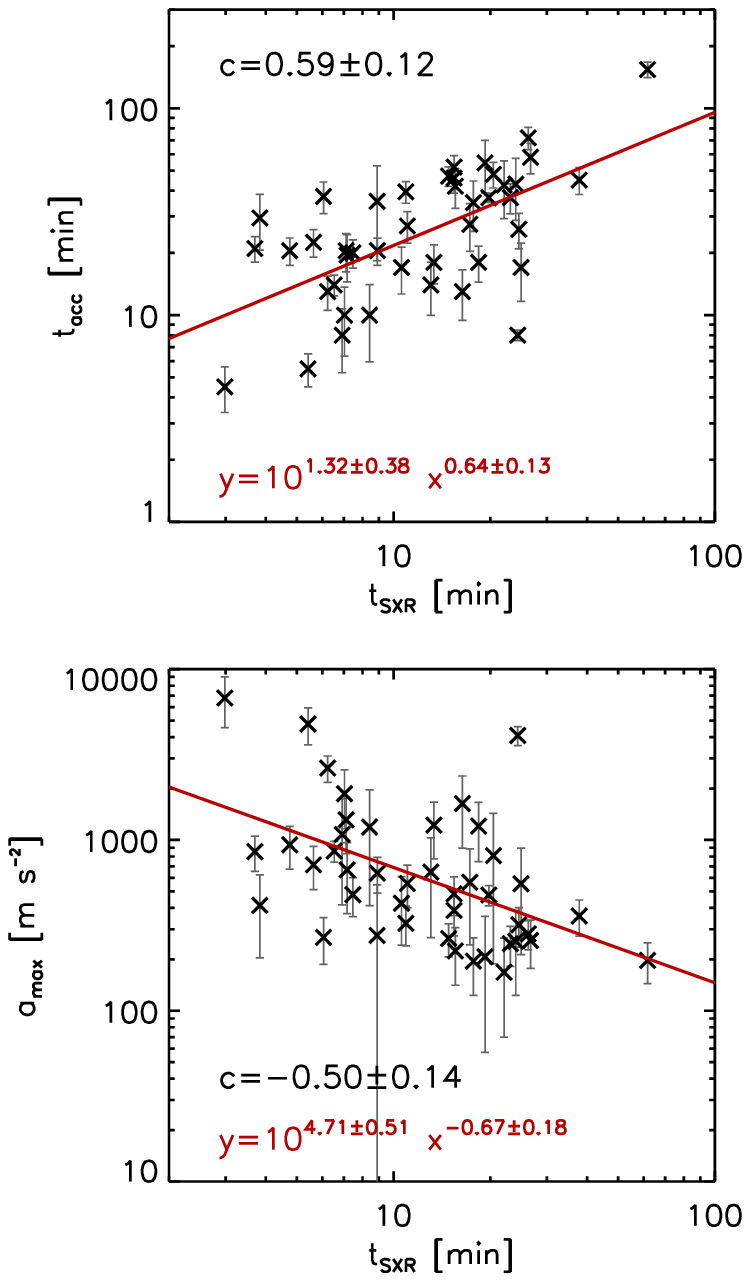}
	\caption{Same as Figure \ref{risetime} but only for events, where the CME acceleration starts before the flare.}
	\label{risetime1}
\end{figure}

For the majority of the events (75\%) we found that the CME acceleration starts \textit{before} the SXR flare onset, which is consistent with the findings of \citet{maricic2007}, suggesting that the flare is a consequence of the eruption. Similar to our study these authors also found about one fourth of the events, for which the onset of the associated flare occurred before the CME acceleration started. They explained these cases by a superposition of two flares, a confined flare in the pre-eruption stage, which releases only a part of the stored magnetic energy, and a second flare, beginning after the CME acceleration onset and causing a prolongation of the first flare in the full-disk integrated SXR light curve. The CME is associated with the second flare but because of the superposition, the flare start is measured from the first one. To test this hypothesis, we checked the SXR curves for all events, for which the CME acceleration start was after the flare onset and found indeed in 11 out of 14 events evidence of a second SXR peak, confirming their suggestion \footnote{This is a significantly higher rate than in the total sample of events, in which about 20\% showed a double SXR peak}. Figure \ref{accsxr} shows two representative examples. The top panels show the CME acceleration profile, the bottom panels the GOES SXR flux together with its derivative. We marked the two possible SXR peaks by arrows. Assuming that we have a superposition of two subsequent flares, we would also measure erroneous SXR start times and as a result too long $t_{SXR}$. This misidentification would also influence our correlation negatively. To test this, we correlated $t_{SXR}$ again with $a_{max}$ and $t_{acc}$ considering only events for which the CME acceleration starts before the flare onset and found indeed distinctly higher correlation coefficients (Figure \ref{risetime1}, $c=0.59\pm0.12$ and $c=-0.50\pm0.14$) than for the whole sample (shown in Figure \ref{risetime}, $c$=0.37$\pm$0.15 and $c=-0.32\pm0.15$). Thus superposition of two subsequent flares and more complex structures are probably a reason for the weak correlations and may account for a considerable number of events, where the flare seems to start before the eruption. 

For the majority of the events (77\%) we found that the end of the CME acceleration occurred \textit{after} the SXR peak. Especially long duration flares reach a certain point, when they become too weak to compensate cooling of the hot plasma. As a result the SXR curve decreases although the energy release in the flare may still be going on.

For 81\% of the events the time delay between the CME acceleration peak and the peak of the GOES SXR flux derivative, which is a proxy for the flare energy release rate, was smaller than $\pm$10 min, for 58\% smaller than $\pm$5 min. This high synchronization hints at a feed-back relationship between the CME and the flare energy release \citep{lin2004, zhang2006, maricic2007, temmer2008, reeves2010, temmer2010}. There are basically two forces acting on a flux rope in equilibrium, an upward directed magnetic pressure and a downward directed magnetic tension of the overlying magnetic field. When the magnetic structure looses equilibrium, it starts rising and a current sheet is formed below it, where magnetic reconnection takes place \citep{priest2002}. The reconnection reduces the tension of the overlying field and enhances the magnetic pressure at the bottom part of the flux rope due to additional poloidal flux, providing the upward acceleration of the rope \citep{vrsnak2008}. The upward motion of the rope leads to elongation of the current sheet and a more efficient reconnection, thus enhancing the acceleration. On the other hand, more efficient reconnection means also a more powerful energy release in the CME-associated flare, which directly relates the dynamics of the eruption and the energy release in the flare.

\acknowledgements
This work was supported by the \"Osterreichische F\"orderungsgesellschaft (FFG) of the Austrian Space Applications Programme (ASAP) under grant no. 819664 and by the Fonds zur F\"orderung wissenschaftlicher Forschung (FWF): P20867-N16 and V195-N16. 
The STEREO/SECCHI data are produced by an international consortium of the Naval Research Laboratory (USA), Lockheed Martin Solar and Astrophysics Lab (USA), NASA Goddard Space Flight Center (USA), Rutherford Appleton Laboratory (UK), University of Birmingham (UK), Max-Planck-Institut f\"ur Sonnensystemforschung (Germany), Centre Spatiale de Li\`{e}ge (Belgium), Institut d'Optique Th\'{e}orique et Appliqu\'{e}e (France), and Institut d'Astrophysique Spatiale (France).

\bibliographystyle{aa}

\end{document}